\documentclass[twocolumn, final]{svjour3} 
\usepackage{mathptmx}      
\usepackage{amsmath}
\usepackage[sort&compress,round,comma,numbers]{natbib}
\usepackage{marvosym, pifont}
\usepackage{graphicx}
\usepackage[colorlinks,
            citecolor=blue,
            urlcolor=blue,
            bookmarks=false,
            hypertexnames=true,
            linkcolor=blue]{hyperref} 

\def\makeheadbox{{%
\hbox to0pt{\vbox{\baselineskip=10dd\hrule\hbox
to\hsize{\vrule\kern3pt\vbox{\kern3pt
\hbox{\bfseries Experiments in fluids}
\hbox{This is a pre-peer review, pre-print version of this article.}
\hbox{The final authenticated version is available online at: \href{https://doi.org/10.1007/s00348-020-03121-3}{https://doi.org/10.1007/s00348-020-03121-3}.}
\kern3pt}\hfil\kern3pt\vrule}\hrule}%
\hss}}}
\renewcommand\vec{\mathbf}

\journalname{Experiments in Fluids}

\begin{document}

\title{Design, construction and validation of an instrumented particle for
the lagrangian characterization of flows}

\subtitle{Application to gravity wave turbulence}

\titlerunning{}        

\institute{ \and Departamento de F\'isica, Facultad de Ciencias Exactas y
Naturales, Universidad de Buenos Aires, Ciudad Universitaria, Buenos Aires 1428,
Argentina; IFIBA, CONICET, Ciudad Universitaria, Buenos Aires 1428, Argentina.}

\author{Facundo Cabrera \and Pablo J. Cobelli}

\authorrunning{F. Cabrera \and P.~J. Cobelli} 

\institute{%
    F. Cabrera \at
        Univ Lyon, ENS de Lyon, Univ Lyon 1, CNRS, \\
        Laboratoire de Physique, F-69342 Lyon, France \\
        \email{facundo.cabrera@ens-lyon.fr}           
\and
    P. J. Cobelli \at
        Departamento de F\'\i sica, Facultad de Ciencias Exactas y Naturales \\
        Universidad de Buenos Aires \& IFIBA, CONICET, \\
        Ciudad Universitaria, Buenos Aires 1428, Argentina \\
        \email{cobelli@df.uba.ar}           
}

\date{Received: date / Accepted: date}

\maketitle


\begin{abstract}
The design and application of an instrumented particle for the lagrangian
characterization of turbulent free-surface flows is presented in this study.
This instrumented particle constitutes a local measurement device capable of
measuring both its instantaneous 3D translational acceleration and angular
velocity components, as well as recording them on an embarked removeable memory
card. A lithium-ion-polymer battery provides the instrumented particle with up
to 8 hours of autonomous operation.  Entirely composed of commercial
off-the-shelf electronic components, it features accelerometer and gyroscope
sensors with a resolution of 16~bits for each individual axis, and maximum data
acquisition rates of 1 and 8 kHz, respectively, as well as several
user-programmable dynamic ranges. Its ABS 3D-printed body takes the form of a
36~mm diameter hollow sphere, and has a total mass of $(19.6 \pm 0.5)$~g, 
resulting in an effective mass density of 819~kg/m$^3$, which can be adjusted by 
incorporating ballasts inside its body. Controlled experiments, 
carried out to calibrate and validate its performance
showed good agreement when compared to reference techniques.

In order to assess the practicality of the instrumented particle, we apply it 
to the statistical characterization of floater dynamics in experiments of
surface wave turbulence. In this feasibility study, we focused our 
attention on the distribution of acceleration and angular velocity fluctuations
as a function of the forcing intensity. The IP's motion is also simultaneously
registered by a 3D particle tracking velocimetry (PTV) system, for the purposes
of comparison. Our results show, among other findings, that the IP is able to 
register the gradual departure from gaussianity and the emergence of strongly 
non-gaussian distributions for the translational acceleration fluctuations as 
the forcing amplitude is progressively increased. 

Beyond the results particular to this study case, the quality of their agreement
with their PTV counterparts and with previous works reported in the literature,
constitute a proof of both the feasibility and potentiality of the IP as a tool for 
the experimental characterization of particle dynamics in such flows.
\\ 
\keywords{measurement techniques \and free surface \and wave turbulence \and 
    inertial particles }
\PACS{PACS code1 \and PACS code2 \and more}
\end{abstract}



\section{Introduction}

In the last two decades, considerable progress has been made in the development
of experimental techniques to measure particle transport properties of flows
\cite{annual_rev_imaging,Virant_1997}. Most experimental techniques commonly
used in the characterization of transport processes in turbulent flows (as is
the case of, e.g., particle tracking velocimetry) are based on high speed
cameras tracking the trajectory of particles in the flow
\cite{nature_dudderar_PIV,Trucco,MTV}.  Often, ultra-fast cameras are necessary
to extract particle translational accelerations using optical techniques, so as
to be able to take multiple derivatives of a registered trajectory at discrete
times. Measurement of angular velocities, in turn, also presents considerable
experimental challenges \cite{25,roco,24,bordoloi_variano_2017,ellen_rotation}. 

Other families of techniques involve the use of wave sources interacting with
seeding particles in the flow. This is the case, e.g., of both imaging laser Doppler
velocimetry \cite{Imaging_laser_Doppler_velocimetry} and ultrasound imaging
velocimetry \cite{Ultrasound_Imaging_Velocimetry}. In contrast, techniques 
relying on marking lines of fluid particles, such as flow tagging based on the
vibrational excitation of oxygen in air flows \cite{Miles87,Miles89} and
molecular tagging velocimetry (MTV) in the case of gas and liquid phase flows
\cite{koo1993,koo1997,koo1999}, do not require seeding. 

In either case, the implementation of such optic- or sonic-based measurement
techniques to characterize flows is impractical in a number of cases,
particularly in the situations where the flow is not directly accessible or when
the region of interest is large. Such situations are widely present both in the
industry and in nature. Some examples of the latter are litter, debris or
clusters of algae floating on a large body of water. Industrial applications
range from mixing tanks to flows in opaque pipes. In all these scenarios, being
able to assess the statistics of particles in the flow represents a substantial
advantage, providing valuable information on the transport dynamics of the flow.

For these reasons, alternative measurement techniques involving the use of
particles with embedded sensors have been proposed. One of the first
implementations of such strategy can probably be traced back to the use of a
thermo-sensitive liquid crystal particle in the investigation of heat transfer
in fluid flows (see \cite{liquidcrystalreview}, and references therein).  In
recent years, \cite{gaste2007b} employed a particle with a temperature
sensor to measure the lagrangian temperature in a Rayleigh-B\'enard flow. Later
on, \cite{smartparticle1,smartparticle2} developed a similar device
capable of measuring both its own 3D acceleration and temperature. Based on
their experimental results, characteristics of the ambient turbulent flow were
inferred from the probability density functions of the particle's translational
acceleration and angular velocity \cite{bodenschatz1,mica1,XU20082095}.  

In this work we propose the utilization of an enhanced instrumented particle
(herein, IP) to characterize particle dynamics in complex laboratory flows.
Generally speaking, this IP takes the form of a centimetric-size sphere
embarking a 3-axis accelerometer and 3-axis gyroscope, as well as a solid-state
storage device; all controlled by a microcontroller unit and powered by a
lightweight rechargeable battery. This allows for on-board measuring and
recording of the IP's own translational acceleration and angular velocity.
Moreover, the density of the IP can be adjusted to have positive or negative
buoyancy.

Being able to simultaneously measure its own acceleration and angular velocity,
(closely related to forces and torques) renders the IP an valuable tool for
experimental research on the dynamics of inertial particles in turbulent flows,
where such quantities play a central role in the development of theoretical 
models \cite{gasteuil,Bellani421973,romain}. 

In what regards to the particularities of the instrumented particle developed 
in this study, several aspects are worth highlighting. The first one concerns
the electronics embarked in the device. In this respect, the choice was made to
build it from commercial off-the-shelf (COTS) components, in order to ensure 
accessibility and reproducibility, while benefiting from low production costs 
and open documentation. Indeed, since the advent of the Internet of Things (IoT),
a wide variety of sensors based on microelectromechanical systems (MEMS) of 
relatively high precision and performance have been made available to the
public, as is also the case for rechargeable high-capacity power supplies 
of low mass and reduced volume. As a by-product of this design choice, the 
inclusion of additional sensors (such as magnetic field, pressure, salinity) 
is readily available and possible with minor modifications. A second aspect to
mention, when compared to preceding developments, is the addition of the gyroscope,
which enables the calculation of the non-inertial acceleration terms. Moreover,
as the data is stored in a memory card inside the IP, there is no need for 
external devices acting as data receivers of the particle's readings. On the
other hand, this last feature presents two shortcomings. 
Including the data logger board imposes a restriction on the minimum diameter 
of the particle, which in turn introduces a limitation in the scales of the flow 
the particle is able to resolve. Moreover, as writing data to a non-volatile
memory device is a power-intensive operation, the overall power consumption of
the system is increased, requiring a larger capacity battery to maintain the
same level of autonomy. Secondly, an embarked memory implies that the user
must be able to recover the particle after the measurement took place in order 
to access the data. This might not always be the case, and for this reason the
range of applicability of this instrument is mostly limited to laboratory 
experiments where this is possible.

The structure of the paper is as follows. Section 2 describes the instrumented
particle developed for this study, including its electronic components and physical
properties. Section 3 regards calibration of the sensors, and Section 4 concerns 
their validation. An application of the instrumented particle to the
characterization of surface wave turbulence is the subject of Section 5.
Finally, Section 6 presents our conclusions.


\section{Instrumented particle}

\subsection{Electronic components}


The key element of the instrumented particle is the MPU-6050 board (InvenSense
Inc.) which integrates, into a single chip, two independently controlled MEMS
devices: a 3-axis accelerometer and 3-axis gyroscope \cite{mpu6050datasheet}.
The accelerometer employs separate proof masses for each axis. The mass
displacement produced by acceleration along any given axis is detected by
capacitive sensors.  Three independent vibratory gyroscopes detect rotation
about each of the three mutually perpendicular axis. Rotation about an axis
induces a vibration (via the Coriolis effect) that is detected by a capacitive
pickoff. Both accelerometer and gyroscope signals are amplified and filtered  to
produce a voltage proportional to the corresponding magnitude. These analog
signals are then digitized using six individual on-chip 16-bit ADCs (one for
each sensor output channel: 3-axis accelerometer and 3-axis gyroscope). This
also enables simultaneous sampling of each signal while requiring no external
multiplexer. 

Dynamic ranges for the accelerometer and the gyroscope are independently
programmable from presets. For the former these ranges are $\pm 2$,  $\pm 4$, $\pm
8$, and $\pm 16$~g$_0$%
\footnote{Herein, we employ `g$_0$' to denote the local 
    acceleration due to gravity, so that the symbol `g' is reserved to 
    unambiguously denote the gram.
}%
; while for the latter the available ranges are $\pm
250, \pm 500, \pm 1000$, and $\pm 2000 \: ^\circ$/s (i.e., degrees per second),
corresponding approximately to $\pm 4.36, \pm 8.73, \pm 17.45,$ and $\pm
34.91$~rad/s%
\footnote{
Rotation rate will be specified in rad/s, except when referring to the gyroscope
dynamic ranges, in which case $^\circ$/s will be used in order match
the device documentation \cite{mpu6050datasheet}.%
}.  For the accelerometer, the maximum output data rate is 1~kHz, whereas the
gyroscope maximum sampling frequency is 8~kHz.

In what regards to power consumption, it is worth mentioning that the
accelerometer normal operating current is $500 \mu$A, five times lower than that
of the gyroscope, rated at 3.6~mA. The MPU-6050 operates with a DC power supply
voltage in the range 3.3-5 V.

The results of our preliminary characterization showed that sensing noise
for each component of both accelerometer and gyroscope is gaussian distributed,
irrespective of the dynamic range. Table~\ref{tab:noise} presents the noise
standard deviation values for the dynamic ranges $\pm 16$~g$_0$ and $\pm
2000$~$^\circ$/s; where $a_i$ and $\omega_i$ represent the translational
acceleration and angular velocity, respectively, for the $i$-axis 
of the corresponding sensor.

Data from the sensors is recorded to a non-volatile memory on-board the
particle, by means of an OpenLog data logger (SparkFun Electronics).
This board, controlled by its own 16~MHz ATmega328 microprocessor (Atmel
Corporation), enables data registering to a removable memory card. It comes
preloaded with the Optiboot bootloader which makes it compatible with the
Arduino Uno board. This is particularly useful for programming it through the
Arduino IDE.

This data logger works over a serial connection and supports standard microSD
FAT 16/32 memory cards from 512 Mb up to 32 Gb in storage capacity. Data
transfer baud rates are configurable, up to a maximum of 115,200 bps.  Our tests
showed that the OpenLog draws approximately 5 mA in idle mode and up to 45 mA
while recording data to the memory card; moreover, these values vary depending
on the size of the microSD card and the baud rate. Lastly, a minimum voltage input of
3.3~V is required to operate the data logger. 

\begin{table}[t!]
\centering
\begin{tabular}{lccc}
    \hline
\\[-0.75em]
  Sensor Noise Std Dev  & $\hat{x}$          & $\hat{y}$            & $\hat{z}$   \\
\\[-0.75em]
  \hline
\\[-0.75em]
$\sigma(a_i)$ \: [g$_0$] & $6.6 \times 10^{-3}$ & $6.7 \times 10^{-3}$ & $7.0 \times 10^{-3}$ \\
\\[-0.75em]
$\sigma(\omega_i)$ \  [rad/s] & $2.6 \times 10^{-2}$ & $2.6 \times 10^{-2}$ & $4.9 \times 10^{-2}$ \\
\\[-0.75em]
    \hline
\end{tabular}
\caption{Accelerometer and gyroscope noise standard deviation in units of g$_0$
and rad/s (respectively) for each of the three axes of the sensors,
corresponding to the $\pm 16$~g$_0$ and $\pm 2000$~$^\circ$/s dynamic ranges.}
\label{tab:noise} 
\end{table}

In order to automatically control the IP data acquisition and recording, an
ATtiny85 chip (Atmel Corporation) is employed as the central processing unit of
the IP. This is a 8-bit high-performance, low power
AVR\textsuperscript\textregistered \:   CMOS microcontroller with 8 kb in-system
programmable flash memory, and a maximum clock speed of 20 MHz
\cite{attiny85manual}. Its power consumption when powered with 3.7 V and
operating at the standard 8 MHz frequency (provided by its internal resonator)
is approximately 5 mA. 

Communication between the microcontroller and its peripherals, namely the
sensors and the data logger, is accomplished as follows. The ATtiny85
microcontroller reads the raw data stemming from the MPU-6050 sensors via the
I$^2$C (inter-integrated circuit) communication protocol, and passes those
values, as well as an associated timestamp, to the OpenLog via a serial
connection. The data logger continously records that incoming data stream into a
file on the memory card. 

Power is supplied by means of a 3.7~V, 150~mAh rechargeable lithium-ion polymer
(LiPo) battery model HW651723P. Among those providing sufficient power for this
application, this model was chosen due to its low weight (5 g) and relatively 
small volume ($25 \times 16 \times 6.5$)~mm$^3$. Once fully charged, this power 
source endows the IP with up to 8 hours of continuous operation.

\begin{figure}[t]
\centering
  \centering
  \includegraphics[trim=35 140 190 80,clip, width=\linewidth]{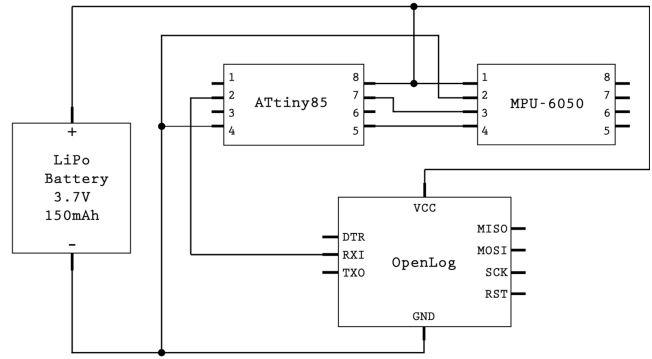}
  \caption{Schematic diagram of the IP's electronic circuit. The ATtiny85
      microcontroller is employed to control both the MPU-6050 sensors (top right) and the OpenLog 
      data logger (bottom right). All three elements are powered by the LiPo battery
      (to the left of the diagram). Connection points between wires are marked
  with small black circles.} 
        \label{fig:conection}
\end{figure}

The electronic connections of the instrumented particle components are presented
schematically in Figure~\ref{fig:conection}. 

\subsection{Physical properties of the instrumented particle} 

For the body of the instrumented particle a spherical casing was designed
entirely on a 3D CAD parametric modelling software (SOLIDWORKS, Dassault
Syst\`emes).  This casing is composed of two hollow spherical caps resulting
from cutting a spherical shell along a chord, as seen on Figure~2. The inner
side of each cap is designed to hold the IP's electronic components in
predetermined fixed positions (relative to the casing) aimed at homogeneizing
the mass distribution inside the IP. Mechanical joining of the two parts is
achieved through an internal single-turn equatorial thread. The shell is
3D-printed in an ABS-based production-grade thermoplastic (ABSplus-P430), 
by means of a Stratasys uPrint SE printer with a layer
resolution of 0.254~mm.  According to the manufacturer's specifications, the
specific gravity of the material is 1.04.  

Within the spherical casing, the sensors are purposedly located at the
geometrical center of the particle, which is then made to coincide with its
center of mass (to within $\pm 0.5$~mm uncertainty) by means of the addition of
appropriately placed ballasts. 

The external diameter of the IP is $D_{e} = (36.0 \pm 0.1)$~mm; this rather
large size is imposed by the dimensions of the electronic component boards and
the battery. The total mass of the instrumented particle, determined
experimentally, is $m_\text{IP} = (19.6 \pm 0.5)$~g; in agreement with the
estimated 19.41~g arising from the technical specifications of the electronic
components and the mass of the shell as calculated via the CAD software 
(the wires and electric connections, not included in the CAD model, might
be the origin of this slight difference). This
mass leads to an effective mass density of $\rho_{IP} = 819$~kg/m$^3$ for the
particle, which can be increased either by modifying the inner geometry of the
3D-printed part or simply by incorporating ballasts inside the shell (e.g., in
the form of patches of tungsten conductor paste). 

\begin{figure}[t]
\centering
\includegraphics[trim=40 0 50 0, clip, scale=1.1]{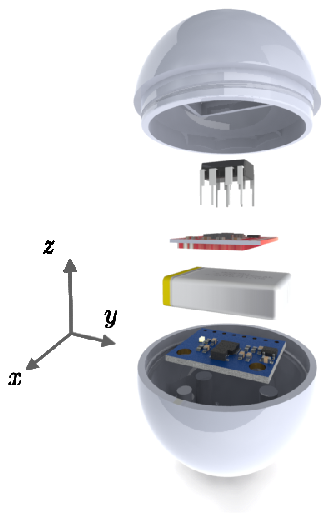}%
\includegraphics[trim=50 0 0 100, clip, width=0.27\textwidth]{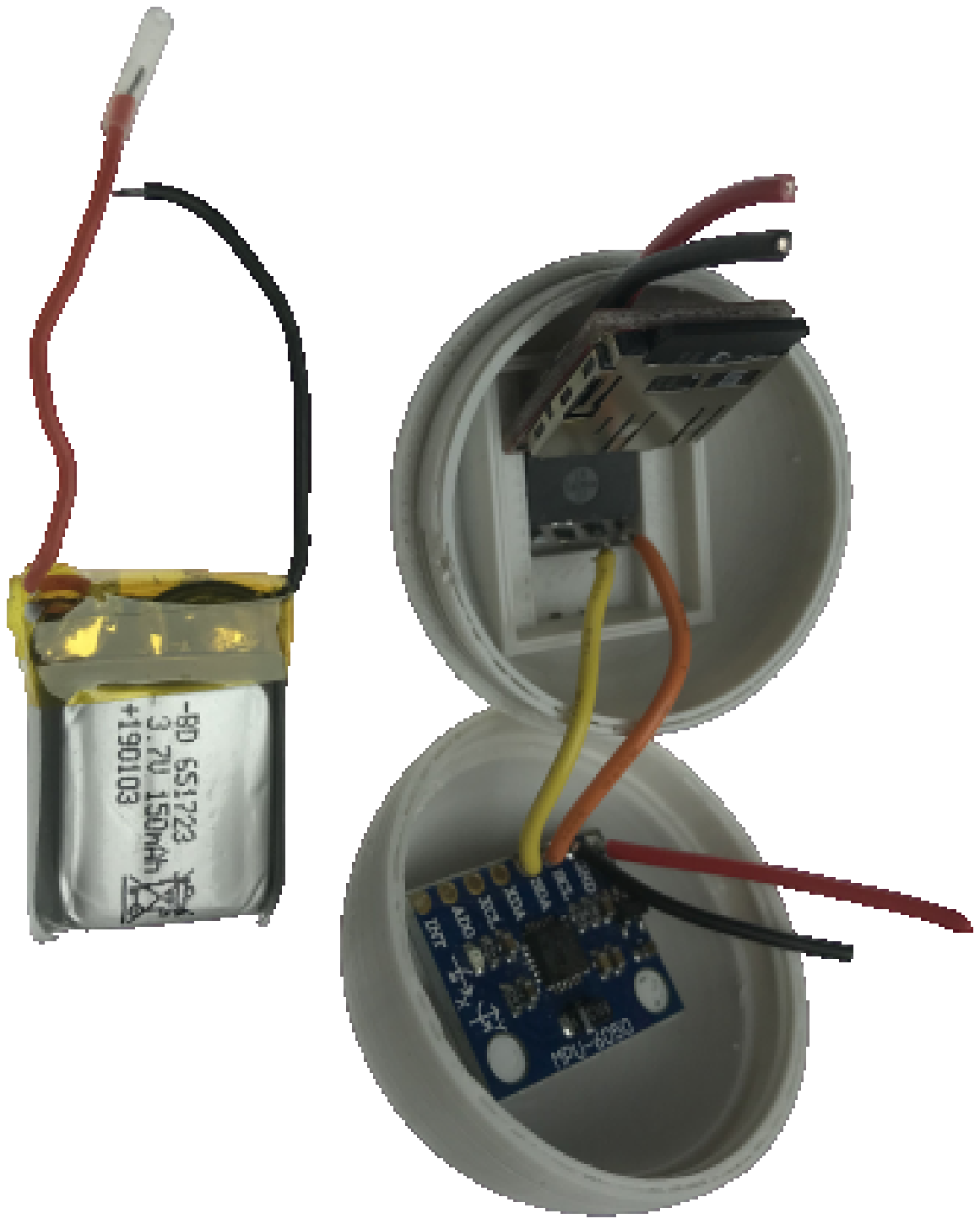}
\caption{(Left panel) Photorealistic render of the instrumented particle's
    components and their internal arrangement. Surrounded by the hollow
    spherical caps, this diagram shows, from bottom to top: the MPU-6050
    (in blue inside the casing), the LiPo battery, the Openlog (in red), and
    the ATtiny85 chip. The inner sides of the caps feature custom designed slots
    to keep the electronic components from moving.(Right panel) Photograph 
of the instrumented particle, prior to final assembly. }
\label{fig:solid}
\end{figure}

The left panel of Figure~\ref{fig:solid} shows a photorealistic render of the
instrumented particle developed in this work, displaying both the IP's
components as well as their internal arrangement. Upon closing the IP by
screwing the caps together, the equatorial inner thread forces a 90$^\circ$
rotation on the OpenLog, the plugged memory card, and the ATtiny85, 
relative to the position shown in this exploded view. This design detail is 
intentionally introduced to minimize the bias that aligning all electronic 
components (with their flat and elongated shape) could have on the 
moments of inertia of the particle. In this schematic representation, 
connections between components are not shown for the sake of visual simplicity. 
Complementarily, the right panel of Figure~\ref{fig:solid} depicts a photograph
of the instrumented particle prior to final assembly. In this picture, the 
inner geometry of the caps is clearly visible, showing the slots designed to 
keep the electronic components fixed at specific positions inside the particle.

From the 3D CAD design software it is possible to accurately calculate the
principal axis of rotation as well as the principal moments of inertia of a
given mechanical assembly, provided realistic mass distributions for the
individual components are supplied. For the IP, the principal axes of rotation
(defined relative to its center of mass) coincide with the accelerometer and
gyroscope axes (to within 1\% uncertainty). Moreover, the estimated values of
the principal moments of inertia of the particle are $J_x = 1969$~g mm$^2$ and
$J_y = 1996$~g~mm$^2$, for the directions lying on the plane of the equator, and
$J_z = 1897$~g mm$^2$ for the perpendicular. The first two differ in less than
1.5\%, whereas $J_z$ is smaller than its counterparts by a 5\% factor. 

After the particle is closed and prior to immersing it in the flow, the IP is
sprayed uniformly with a peelable rubber coating (Rust-Oleum Peel Coat). This
covers the exterior of the particle with a black film of thickness 0.1~mm,
providing a waterproof seal for the IP for a minimum lapse of 6 hours. This seal
is removeable once the measuring campaign is completed and the particle is
opened to recover the memory card holding the acquired data.

Digital design files and Arduino codes are both available upon request 
from the authors.

\subsection{Particle dynamics and sensors' signals}

Due to the IP rotation and the presence of gravity, the acceleration measured by
the instrumented particle is not exactly its translational acceleration. In this
sense, a time dependent rotation matrix, which represents the instantaneous
rotation between laboratory and particle reference frames, is considered.
Additionally, centrifugal, Euler and gravity components of acceleration have to
be taken into account (note that the Coriolis contribution is null, however, due
to the sensor being fixed within the sphere). Hence, the relation between the
translational acceleration $\vec{a}_{trans}$ and the acceleration $\vec{a}_{IP}$
measured by the IP is given by
\begin{equation}
\underline{\underline{\vec{R}}}(\vec{\alpha}(t))~\vec{a}_{IP}= \vec{a}_{trans} 
+ \vec{\omega}\times\big(\vec{\omega}\times \vec{r}\big) 
+ \vec{\dot{\omega}}\times\vec{r} + \vec{g}_0,
\label{eq:acceleration}
\end{equation}
where $\vec{\omega}$ and $\dot{\vec{\omega}}$ are the angular velocity and
acceleration, respectively, and $\vec{r}$ is the vector defined by the geometric
center of the particle and the sensor position. Besides,
$\underline{\underline{\vec{R}}}(\vec{\alpha}(t))$ is the 3D rotation matrix
that relates the particle and laboratory reference frames and it depends on the
3D angular position of the particle, $\alpha (t)$. Notice that in the case 
in which the non-inertial components of the acceleration are negligible and
the rotation is isotropic, it is straightforward to relate $a_{trans}$ with
$a_{ip}$; whereas in the opposite case it is necessary to make use of the 
angular velocity measurements to derive those contributions. In either case,
it is possible to calculate even central moments of $a_{trans}$  
\cite{ZimmermannThesis}.


\section{Calibration}

The 3D accelerometer requires a calibration to obtain its offset and
sensitivity. The procedure of calibration is based on the fact that if the
sensor is at rest, the norm of the measurements must be equal to the
local acceleration of gravity:%
\begin{equation} 
\left( \frac{A_x-O_x}{S_x} \right)^2 + \left(\frac{A_y-O_y}{S_y}\right)^2
    + \left(\frac{A_z-O_z}{S_z}\right)^2  = \text{g}_0^2,  
    \label{eq:norm_acceleration}
\end{equation} 
where the vectors $\vec{O} = (O_x, O_y, O_z)$ and $\vec{S} = (S_x, S_y, S_z)$ 
are the sensor's offset and sensitivity;
respectively, and the raw accelerometer readings in each
direction are represented by $\vec{A} = (A_x, A_y, A_z)$.
In order to obtain
$\vec{O}$ and $\vec{S}$ from this equation, the following procedure is carried
out. Time series of raw 3D-acceleration data are acquired for 9 mutually 
independent orientations of the sensor (parallel and antiparallel to gravity
for the three proper axes, plus three arbitrary orientations) while at rest.
For each instantaneous determination of the 9 raw acceleration time-series, 
eq.~\eqref{eq:norm_acceleration} results in a system of equations that can 
be solved to obtain the values of sensitivity and offset for each axis. 
In all cases, we employed time-series data consisting of 10,000 data points 
(corresponding to a 10~s acquisition lapse at the maximum sampling frequency)
for each orientation; obtaining an equal number of determinations of $\vec{O}$
and $\vec{S}$.

\begin{figure}[t!]
\centering
\includegraphics[width=1\linewidth]{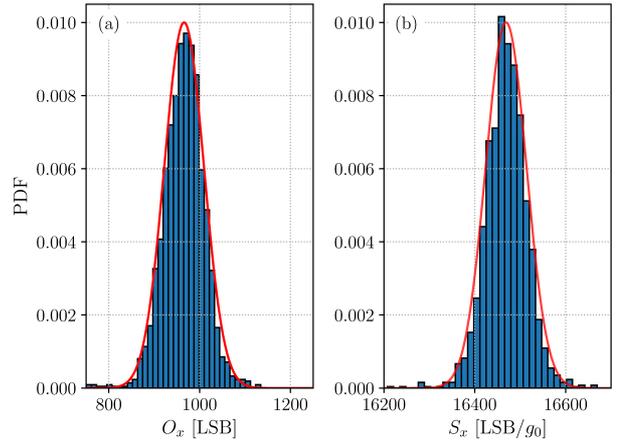}
\caption{Distributions for offset (left panel) and sensitivity (right panel)
    derived from the accelerometer calibration described 
    in the text. Results are only shown here for the $x$-axis and $\pm 2$~g$_0$; 
    qualitatively similar behaviors were obtained for other axes and dynamic ranges.
    In both panels, a fit by a gaussian distribution is represented by a (red)
    continuous line. }
\label{fig:offsetysensibilidad}
\end{figure}

Our results show that both offset and sensitivity are gaussian distributed;
we therefore define their values as those resulting from their distribution 
means. As an illustration, panels (a) and (b) of Figure~\ref{fig:offsetysensibilidad}
present these calibration results for the offset and sensitivity distributions
for one particular axis of the MPU-6050 accelerometer operating at the $\pm
2$~g$_0$ dynamic range, respectively. Note that the mean sensitivity value obtained 
is comparable with an estimate based on the ratio of the maximum 
representable integer at the board's bit resolution
and the working dynamic range; namely: 
$S = 2^{16}/4$~g$_0$ = 16,384 in units of LSB/g$_0$, where LSB stands for Least
Significant Bit.

As the values for offset and sensitivity depend also on the 
overall dynamic range chosen for the sensor, this calibration procedure 
was repeated for each of them. Qualitatively similar behaviors were 
observed for other axes and dynamic ranges. 

Finally, in the case of the gyroscope, the corresponding values for offset and
sensitivity were taken from the MPU-6050 datasheet and successfully validated in
the following Section. 


\section{Validation}

\subsection{Gyroscope validation}

The gyroscope performance is validated according to the following procedure.
The particle is mounted at the end of a rotating DC motor's shaft, making sure
the selected sensor axis is properly aligned with it. This can accomplished by
setting the motor in motion, monitoring the signals read by the remaining two
axes for coherent signals (corresponding to projections of the angular velocity)
and correcting for misalignment. Once the axis is aligned, the motor is made to
rotate at a controlled angular frequency while acquiring a time series signal of
the IP's angular velocity.  This operation is repeated for several rotation
frequencies between $0.3$ and $1.2$~rad/s (limited by the geared motor
employed), and for the three axes of rotation of the gyroscope. 

\begin{figure}[t]
\centering
\includegraphics[width=0.5\textwidth]{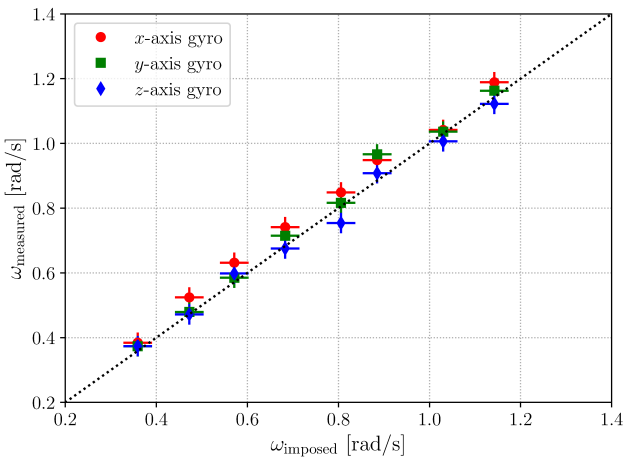}
\caption{Results for the validation of the instrumented particle's gyroscope. 
    Data for the $x$, $y$, and $z$ axis is represented by (red) circles, (green)
    squares, and (blue) diamonds, respectively. A dashed (black) line of slope equal 
    to unity is displayed as a visual reference. }
\label{fig:giroscopo-motor}
\end{figure}

The results of this validation are presented in
Figure~\ref{fig:giroscopo-motor}, for each of the gyroscope axes, along with a
dashed line of slope equal to unity, included as a visual reference. In the
Figure, horizontal error bars are associated with the uncertainty in the imposed
frequency, whereas vertical error bars result from the standard deviation of
each time series. 

Although the specific values for each axis differ slightly, separate linear fits
(not shown in the figure) result in slopes ranging from 0.971 to 1.018, and
absolute values for the corresponding intercepts lower than 0.005 (both
established with an uncertainty of 1.5\%).

Our results are compatible with the values reported in the sensor's
documentation, therefore in what follows we employ those values for the offset
and sensitivity of the gyroscope. 

\begin{figure*}[ht!]
\centering
\includegraphics{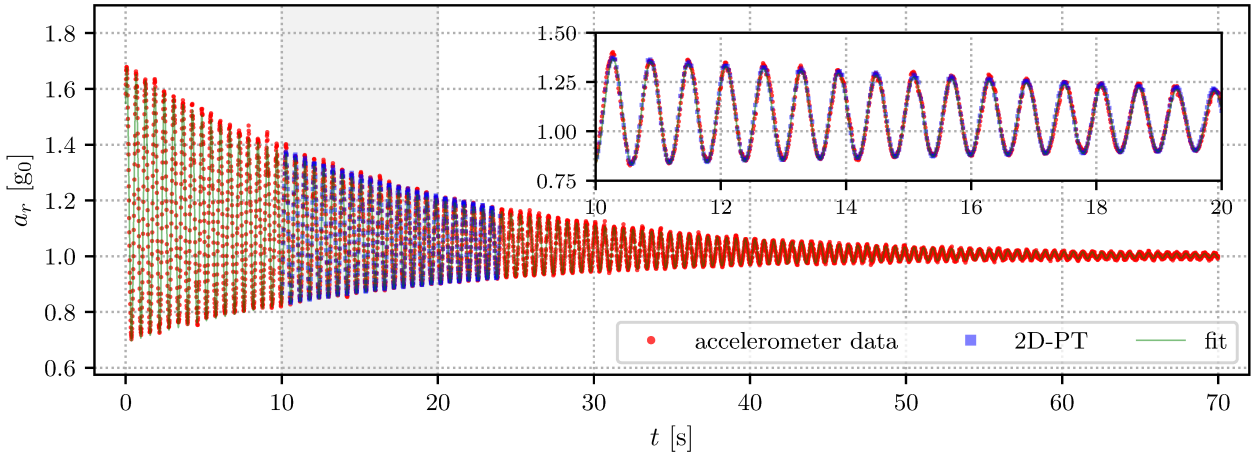}
\caption{Accelerometer validation results through a physical pendulum experiment. 
    Measurements of the IP's sensor radial acceleration, shown in (red) circles, 
    are compared to: (a) its counterparts obtained from PTV, depicted by (blue)
    squares and only available within a reduced time window, and (b) the
    best-fit based on an approximate theoretical prediction, represented by a continuous (green)
line. The inset shows a zoomed view of the three curves in the region
highlighted by a gray background.
}
\label{fig:solapamiento2}
\end{figure*}

\subsection{Accelerometer validation}

For this validation, the sensor is mounted at the tip of a physical compound
pendulum consisting of a thin, stiff rod and a mass attached to its end; capable
of swinging about a pivot point by means of a ball bearing.  During the
oscillating motion, the sensor records its acceleration while a fast camera
simultaneously captures the 2D movement of the pendulum. Particle tracking is
employed later to determine the trajectory of the pendulum, from which the
radial acceleration registered by the sensor can be extracted.

For the acceleration measurements, the accelerometer was set at a sampling
frequency of 100 Hz, and the $\pm 2$~g$_0$ dynamic range was selected.  For
particle tracking, the images were acquired by a Photron Fastcam SA3-60K camera
($1024 \times 1024$~px$^2$ resolution, 12-bits color depth). The image sampling
rate was established at $50$~Hz (lowest available value for this setting on the
camera) and a shutter speed of $1/1000$~s was chosen to minimize the effects of
motion blur on particle tracking. The resulting error in position tracking was
estimated (a posteriori) to be better than 0.5\%. Due to limitations in the
amount of flash memory available, the camera is only able to record for
approximately 25~s during the pendulum motion.

Accelerometer results are also compared to the theoretical prediction
for the radial acceleration as registered by the sensor, given by
\begin{equation}
a_r(t) = r\omega^2 \theta_0^2 e^{-2\gamma t} \sin^2(\omega t) + 
g_0 \cos [\theta_0 e^{-\gamma t } \cos(\omega t)], 
\label{eq:dauphin}
\end{equation}
where $r$ is the distance between the pivot and the position of the sensor,
$\theta_0$ is the initial angle of the motion, $\gamma$ is the damping factor
and $\omega$ is the oscillation frequency, given by $\omega^2 = \omega^2_0 -
\gamma^2$; $\omega^2_0$ being the natural frequency of the undamped pendulum.
It is worth mentioning that equation \eqref{eq:dauphin}, derived in
\cite{articulo_pendulo}, is valid for small damping ($\gamma \ll \omega$) and
under the small angle approximation ($\theta_0 \ll \pi/2$). 

Figure~\ref{fig:solapamiento2} shows the results for the radial acceleration
measured by the sensor (in red circles) for a time lapse of 70~s which covers
the dynamics from the pendulum's release up to its equilibrium state.  These
results are compared to those derived from the 2D particle tracking algorithm
(in blue squares) available for a narrower time window corresponding to $t \in
[10, 24]$~s. A good agreement is observed between the accelerometer data and the
results from particle tracking in this overlapping zone.  

Additionally, parameters of the model equation~\eqref{eq:dauphin} are adjusted
to our accelerometer data; the results from this fit are represented as a
continuous (green) line in the same Figure. The best-fit values obtained for the
parameters are as follows: $\omega = 5.23$~rad/s, comparable to the theoretical
value of 5.8~rad/s based on an estimate for the inertia moment of the pendulum;
$r = 0.37$~m, close to the pendulum's length of 0.39~m; the damping factor value
$\gamma = 0.02$~s$^{-1}$, compatible with a low damping regime; and $\theta_0 =
0.78$~rad, in agreement with the initial angular displacement (of approximately
45$^\circ$) imposed on the pendulum.

Comparison between the accelerometer data 
and the theoretical prediction reveals again good agreement over the entire
motion, both in terms of the frequency of oscillation and the upper and lower
amplitude envelopes. The inset in Figure~\ref{fig:solapamiento2} displays a 
zoomed view of these three signals (accelerometer data, 2D-PT, and theoretical
prediction) in the time interval highlighted with a gray background (where they
coexist), showing a three-fold agreement.

This validation procedure was repeated by changing the axis of the accelerometer
pointing in the radial direction, leading to results similar to those presented
in the preceding paragraphs (omitted in this exposition for the sake of brevity). 


\section{Application to surface wave turbulence}

This section presents the results of the application of a buoyant version of our
IP to characterize a free-surface wave turbulence flow. Typical results
of the statistical properties of acceleration and angular velocity are analyzed.
The goal is to use these results of the IP dynamics in a known flow as a
benchmark for the performance of the instrumented particle and its applicability
to the characterization of flows. 

\subsection{Experimental setup}

\begin{figure}[t]
\centering
\includegraphics[trim={50 0 60 0}, clip, width=1\linewidth]{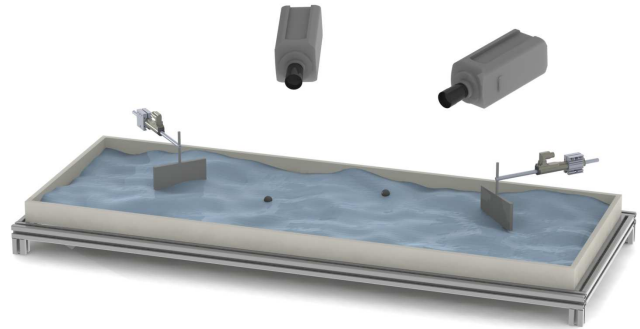}
\caption{Schematic diagram of the experimental setup employed in the study
    of floater dynamics in surface wave turbulence. Waves are created by the
    stirring motion of paddles, independently controlled by servo-motors (see
    text for details on the forcing characteristics). Two identical instrumented
    particles are used, together with a 3D-PTV system composed of two high-speed
    cameras.}
\label{fig:photolab}
\end{figure}

Figure~\ref{fig:photolab} presents a schematic representation of the
experimental setup employed in this Section. The experiments are carried out
in a PMMA wave tank of dimensions (2.0$\times$0.8$\times$ 0.15)~m$^3$. Distilled
water is employed as the working fluid, with the addition of TiO$_2$ powder
in a low concentration (4~g/l) in order to render the liquid
white without significantly changing its rheological properties
\cite{Przadka2012}. This simplifies the detection and subsequent tracking of
the particle for the PTV measurements since a strong color contrast is achieved
between the (black) IP and the surrounding (white) fluid in which it is
partially immersed.

A turbulent steady state of surface waves is generated (and sustained) by
continuously stirring the flow with two partially immersed paddles. The
piston-type motion of these paddles, whose surface area is
(15$\times$10)~cm$^2$, is established by two independently-driven
servo-controlled electromagnetic linear motors (LinMot, model
P01-23$\times$80/210$\times$270) with a peak force of 47~N and a position
accuracy of 0.01~mm. These wavemakers were driven by a random signal with a
white frequency spectrum in the range [0, 4]~Hz (as done in experiments of
gravity wave turbulence; see, for instance, in \citealt{Cobelli2011,delgrosso2019}).
According to this, the signal used for the forcing thus naturally determines a
characteristic time scale of approximately 0.25~s.  As the tank is filled with
liquid up to a height at rest $h_0 = 5$~cm, this forcing leads to waves in the
deep water regime, following a dispersion relation approximately given by
$\omega^2 \approx g_0 k \tanh(k h_0)$, where $k$ represents the wave number.

\begin{figure*}[t!]
\includegraphics[width=\textwidth]{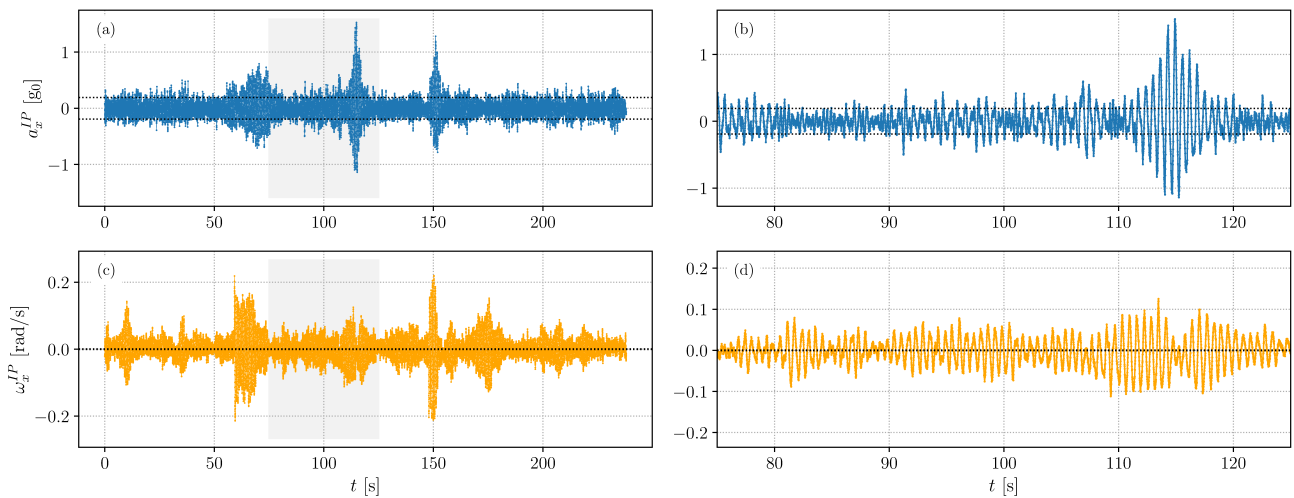}
\caption{Translational acceleration (top row) and angular velocity (bottom row)
    time series as measured by the instrumented particle in the experiments on 
gravity wave turbulence, acquired at a sampling frequency of 100~Hz for 
a forcing amplitude of $A=15$~mm. Only the
$x$ component is shown, qualitatively similar behaviors were registered for the
remaining axes. Panels (b) and (d) show a zoomed view of the time series
depicted in panels (a) and (c), respectively; within the 50~s time window
highlighted in gray. RMS values for $a^{IP}_x(t)$ and $\omega_x^{IP}(t)$ are 
marked in the panels by a dotted (black) line.
}
\label{fig:rawdata}
\end{figure*}

The same maximum amplitude for the wavemakers' motion, symbolized by $A$, was
imposed for to both paddles; its value was then varied for different
experimental runs. For this study, experiments with four different values of the
maximum forcing amplitude $A$ were considered, namely: 5, 10, 15 and 20~mm.
This choice allows for the exploration of regimes with different levels of
nonlinear coupling between waves.  It is worth noting that the amplitude $A$
corresponds to the maximum amplitude for the displacement of the paddle as
measured from its position at rest. Due to the fact that random time signals
were employed for the forcing, the actual mean displacements of the wavemakers
were smaller (e.g., for $A = 10$~mm, the associated standard deviation of the
paddle's position from rest is 2.7~mm). 

In order to track the IP during its motion in the turbulent wave field, a 3D
PTV system was implemented employing two high-speed Photron cameras (a SA-3
camera and a 1024-PCI camera) whose complete field of view covers the entire
wave tank. Calibration results, following the procedure described by
\cite{zhang1999}, resulted in an position uncertainty of 0.3~mm. For both
cameras, the acquisition frequency and shutter speed were set at 60~Hz and
1/250~s, respectively; synchronicity was achieved by means of an external
trigger.  

The experimental protocol is as follows. At first, the wavemakers are set in
motion from an initially rest state. After a transient period (of 3--5~min) in
which a steady state is established, two identical IPs are placed in the
turbulent wave field. For the instrumented particles, the acquisition frequency
for both the embarked accelerometer and gyroscope was set at 100~Hz; whereas the
dynamic ranges chosen for each sensor (based on preliminary tests) are $\pm 4$~g
and $\pm 250^\circ$/s, respectively. Once the cameras' recording time is
elapsed, and during the time period in which images are uploaded to an external
device for storage, the forcing is maintained and the IPs continue to register
data, leading to acceleration and angular velocity time series of a typical
duration of 240~s.

It is worth mentioning that the experimental runs considered for this study do
not include particle collisions (with either the paddles, the wave tank walls or
between each other). Moreover, a posteriori analysis of the PTV results revealed
that, for those selected cases, the distance between the two IPs remained always
greater than 5 (particle) diameters. 

Each experimental run consisted of approximately 40~s of simultaneous IP and PTV
measurement, plus an extended period of 200~s in which only IP data is gathered.
Furthermore, a total of 5 identical (repeated) experiments were carried out for
each value of the maximum forcing amplitude $A$ explored, resulting in a total
of 20 experimental realizations.  

For the particles, 3D data on translational acceleration and angular velocity is
gathered for each experiment.  Typical time series for these magnitudes are
shown in Figure~\ref{fig:rawdata}, corresponding to one particular axis
(arbitrarily chosen, as qualitatively similar results are obtained for other
axes). Panel (a) displays the translational acceleration $a_x$ whereas panel (c)
depicts the angular velocity $\omega_x$, both for the same experimental run,
which corresponds to a forcing amplitude value of $A = 15$~mm. Panels (b) and
(d) display a zoomed view of the corresponding signal in the time window
highlighted in gray (i.e., between 75 and 100~s). RMS values for these signals
are marked in the panels by a dotted black line. Notice that, with peak absolute
values slightly exceeding 1~g$_0$ in the case of the accelerometer and
0.2~rad/s for the gyroscope, both sensors are still far from saturation. 

The accelerometer signals exhibit relatively long lapses of random fluctuations
about a null mean, whose amplitude coincide with the RMS value (which in this
case is 0.19~g$_0$), separated by short-timed intense bursts such as those
occurring at $t \approx 115$~s and $t \approx 150$~s, in which the acceleration
reaches values of order 1~g$_0$. These events point towards the emergence of
sudden and localized variations of the free-surface state at the particle's
instantaneous position, and are associated to dissipative mechanisms in the
flow, such as the formation of cusps and wave breaking. 

A similar behavior is observed in the case of the signals registered by the
gyroscope, as seen on Figure~\ref{fig:rawdata}(c). Moreover, comparison between
panels (a) and (c) shows some degree of correlation between $a_x(t)$ and 
$\omega_x(t)$ for intense events, although this is not always the
case (notice, e.g., the burst in angular velocity occurring at
$t\approx175$~s, with no significant trace in the acceleration time series).  

Finally, both the acceleration and angular velocity time series reveal 
the presence of a carrier frequency of approximately 1.67~Hz, made evident
in the zoomed view of panels (b) and (d). This frequency corresponds 
(roughly) to the center of the frequency band of the forcing signal, 
and therefore appears naturally in the IP signals.

Particle tracking, on the other hand, allows for the determination of
individual particle trajectories from which acceleration can be calculated 
and compared to IP data. In our data post-processing scheme, acceleration
is derived by fitting a series of cubic splines in a piecewise manner to
the discrete positions of the particles at discrete instants in time. This
implies that third order polynomials are fitted between discrete time points,
therefore ensuring continuity of velocity and acceleration. As no 
synchronization between the cameras and the
particles is enforced, the comparison between IP data and results from PTV is
carried out in terms of the statistical properties of their distributions.

Since the instrumented particle has a finite size, the physical scales to which
it is sensitive in the flow under study is naturally limited. Therefore, in
order to draw conclusions on the dynamics observed, it is important to estimate
the minimum size of flow structures (such as waves, in this scenario) the IP 
is able to resolve.   

In order to derive an estimate for this spatial scale, we proceed as follows.
First, the scaling law for the gravity wave turbulence power spectrum
(theoretically predicted and experimentally observed in \citealt{Cobelli2011}) is
considered: $P(f)\propto\omega^{- 17/6}$, where $P$ is the power spectral
density and $\omega$ the (angular) frequency of the waves. The smaller
scales injected in the turbulent wave field are associated to the maximum frequency
of forcing; corresponding, in this case, to $\omega_f=(2\pi\cdot4)$~rad/s, as we
employ forcing frequencies in the 0--4~Hz range. Next, a cutoff frequency whose
power is an order of magnitude less than the power injected into the flow, is
estimated. This cutoff frequency $\omega_c$, obtained from the aforementioned condition given
by $P(\omega_c) = 0.1 \: P(\omega_{f})$, leads to $\omega_{c}\approx 56
$~rad/s. By inverting the dispersion relation for gravity waves, the associated
wave number is calculated, from which a characteristic scale 
can be derived. This is found to be approximately 2~cm, comparable in order 
of magnitude with the size of the
particle. For the angular frequency, the corresponding value for the cut-off is
smaller (by a factor of 2) than the frequency of transition between the gravity
and capillary wave turbulence regimes, as established in \cite{FalconSIAM}. In 
summary, the instrumented particle is sensitive, essentially, to the full range
of surface gravity waves.


\subsection{Angular velocity fluctuations}

\begin{figure}[t]
\centering
\includegraphics[width=0.5\textwidth]{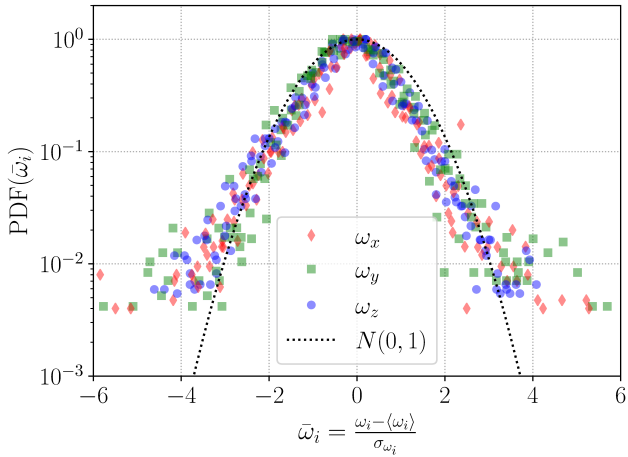}
\caption{Probability density functions of the (normalized) angular velocity 
    registered by the instrumented particle for the largest forcing amplitude 
    considered ($A = 20$~mm). Data is shown 
    with markers: (red) diamonds, (green) squares and (blue) stars denote 
    the $x$, $y$, and $z$ axes, respectively. A standard normal distribution
    is depicted by a (black) dashed line, as a reference. }
\label{fig:histograma_velocidad_angular}
\end{figure}

Angular velocity fluctuations follow a normal distribution, as evidenced from
the associated normalized probability density function (PDF) shown in
Figure~\ref{fig:histograma_velocidad_angular} for the case of the most intense
forcing ($A = 20$~mm). No significant difference between the readings of the
three axes is observed, and qualitatively comparable results are obtained for
the other forcing levels considered in this study. Note that although the tails
of the distribution separate from the normal distribution (depicted by a dotted
black line in the Figure), their point density is significantly lower than that
of the central region. 

In order to quantify the gaussian behavior of the PDFs, a one-sample
Kolmogorov-Smirnov test is performed, testing the null hyphothesis that the
data is drawn from a gaussian distribution (whose mean and standard deviation
are data driven). Our results show that, as anticipated by the PDFs, the 
distributions for the angular velocity fluctuations are indeed gaussian
distributed, with $p$-values below $3 \times 10^{-1}$.

Moreover, from this analysis it is observed that the particles' mean angular
velocities (for all axes and forcing intensities) adopt values of the 
order of $1.6\times 10^{-6}$~rad/s, lower than the noise level of
$5.5 \times 10^{-2}$~rad/s; with RMS values of $4\times10^{-2}$~rad/s that
are above the sensitivity (which, for this dynamic range, is 
$1\times10^{-3}$~rad s$^{-1}$ LSB$^{-1}$).
Peak values for the IP's angular velocity are, in all cases explored, 
lower than 0.26~rad/s.

As can be seen from panels (c) and (d) in Figure~\ref{fig:rawdata}, the particle
rotates about a given axis with alternating positive and negative velocity, 
with a periodicity linked to the characteristics of the underlying flow (recall that the
wave field is generated by forcing at time-scales within the 0--4~Hz band).

Based on these results, and taking into account the 0.5~mm uncertainty in 
the placement of the sensors relative to the particle's center, 
an order-of-magnitude estimate can be given for the non-inertial 
contributions in eq.~\eqref{eq:acceleration} in terms of the maximum angular
velocity and acceleration. According to this, $|\vec{a}_{cen}| \approx 2.2 \times
10^{-4}$~m/s$^2$ and $|\vec{a}_{eul}| \approx 8 \times 10^{-4}$~m/s$^2$. 


\subsection{Translational acceleration fluctuations}

\begin{figure*}[t!]
\centering
\includegraphics[width=\textwidth]{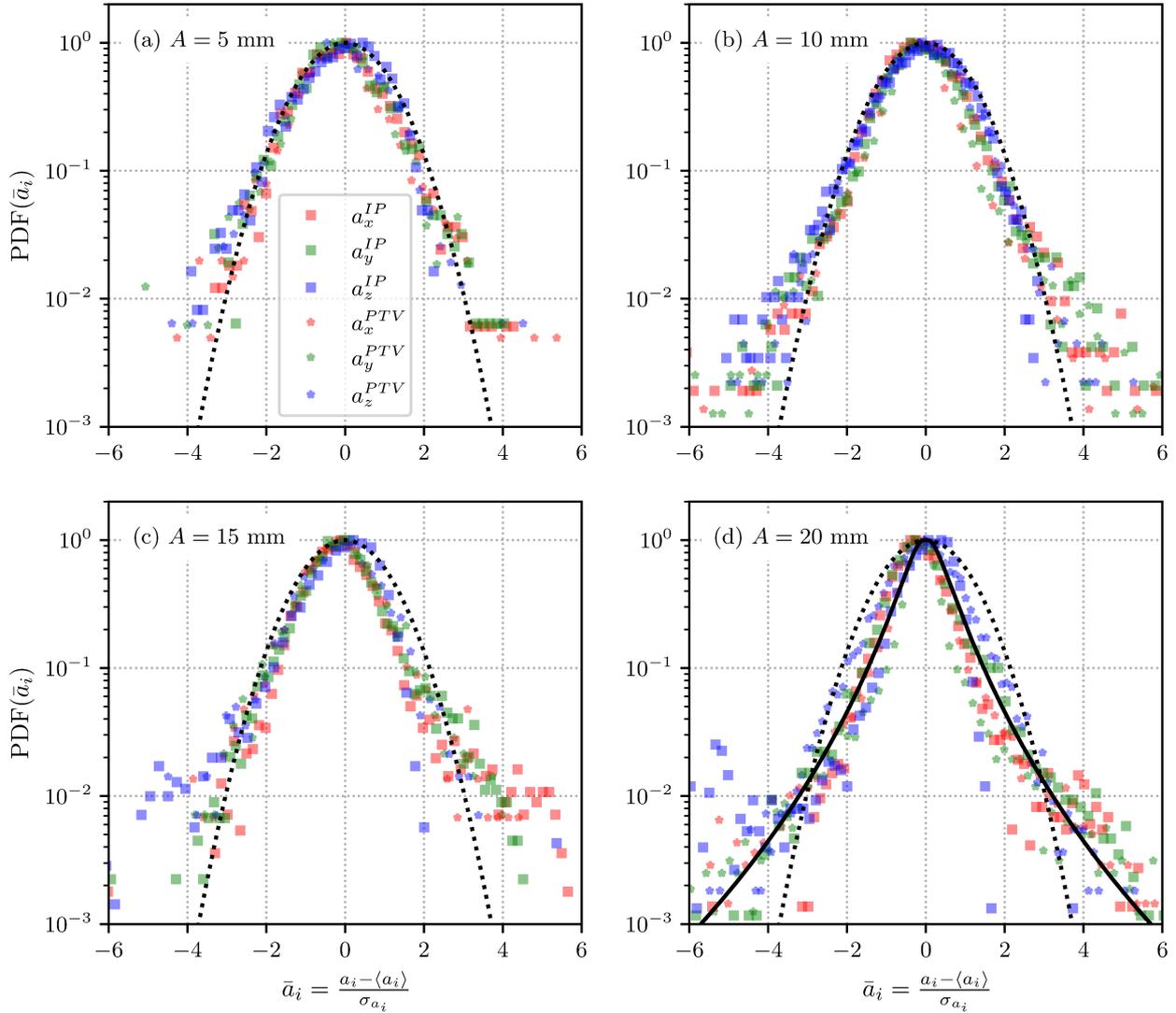}
\caption{Probability density functions of the (normalized) translational
    acceleration of the instrumented particle, for increasing values of the
    forcing amplitude. Data from the instrumented particle and from 3D-PTV
    is represented by squares and stars, respectively. Red, green and blue
    are employed to denote the $x$, $y$, and $z$ axes, in that order. 
    A standard normal distribution is depicted by a (black) dashed line, as a
    reference. Additionally, a stretched exponential 
    distribution corresponding to $\Pi_s(a^{IP}_x)$ as defined by eq.~\eqref{eq:funcion_pi},
    is displayed in panel (d) by a continuous (black) line.}
\label{fig:pdfs_aceleracion}
\end{figure*}

In this subsection the IPs' translational acceleration fluctuations are
considered. Our experimental results for these magnitudes are summarized in
Figure~\ref{fig:pdfs_aceleracion}. Panels (a) through (d) show the normalized
probability density functions for each of the three components of translational
acceleration in increasing order of the forcing amplitude $A$. Results obtained
from the instrumented particles are represented by squares, while those derived
from the particle tracking technique are depicted with a star; each acceleration
component being color-coded so that red, green, and blue denote the $x$, $y$,
and $z$ axes, respectively. For comparison purposes, each panel comprises also a
standard normal distribution, represented by a
dotted black line. 

As seen in the previous subsection, the two non-inertial components of the
acceleration measured by the particle have maximum values below $1\times
10^{-3}$~m/s$^2$, while the measured values of acceleration are of
order 1 g$_0$ (i.e., 9.8~m/s$^2$). Therefore, rotational contributions to 
the acceleration in equation~\eqref{eq:acceleration} can be neglected, 
leading to the following simplified form:%
\begin{equation}
\vec{\underline{\underline{R}}\vec{(\alpha(t))}} \: \vec{a}_{ip}= \vec{a}_{trans} 
    + \vec{g}_0.
\label{eq:acceleration_approx}
\end{equation}

As previously discussed, accelerometer and PTV reference systems are related
by a time dependent transformation hence their axes have
generally different orientations. Due to the fact that the three accelerometer
axes sweep the space homogeneously, we expect the acceleration distribution 
to be similar for each sensor axis, which is compatible to the results shown in
Figure~\ref{fig:pdfs_aceleracion}.

The good agreement found between the results independently obtained from the
instrumented particle and the particle tracking technique contitutes a positive
result indicating the feasibility of application of the IP to characterize this
type of turbulent flows. 

Finally, it is worth considering how the translational acceleration fluctuation
distribution is modified as the forcing is increased. In this sense,
Figure~\ref{fig:pdfs_aceleracion} shows a PDF behavior close to gaussian only
for the lowest forcing (top left panel, corresponding to $A=5$~mm).  As the
forcing is increased beyond that value, the experimental data gradually departs
from the normal distribution and begins to exhibit progressively heavier
exponential tails (and consequently, sharper maxima). This is particularly
evident in the last panel (bottom right, associated to $A=20$~mm, the most
intense forcing considered), where our experimental results show a clearly
non-gaussian behavior. Concurrently, we observe a monotonous growth on the
standard deviation of acceleration fluctuations as the injected power increases.

These results are in agreement with previous experimental studies on other
turbulent flows \cite{11,55,54,72}. In those works, the normalized distribution
for the acceleration $a_i$ along a given $i$-axis is modeled by:%
\begin{equation}
    \Pi_s(a_i) = \frac{e^{3s^2/2}}{4 \sqrt{3}} \left[ 1 - 
    \text{erf} \left( \frac{\ln |a_i/\sqrt{3}| + 2s^2}{s \sqrt{2}} \right)
\right],
    \label{eq:funcion_pi}
\end{equation}
where the value of the parameter $s$ is related to the data kurtosis, $\kappa$,
through the expression:%
\begin{equation}
    \quad s = \frac{1}{2} \sqrt{\ln \left(\frac{5}{9}\: \kappa \right)}.
\end{equation}
The model given by eq.~\eqref{eq:funcion_pi} is extensively employed in the
analysis of intermittency in the dynamics of particles (see, for instance, 
\cite{45} and references therein) and arises from 
from the approximation that the norm of the acceleration presents a
log-normal distribution.

In Figure~\ref{fig:pdfs_aceleracion}(d) the function $\Pi_s (a^{IP}_x)$ is plotted in
a black continuous line, for a value of $s=0.4$ corresponding to a kurtosis of
$\kappa = 3.41$ as determined experimentally from the instrumented particle
data. It is observed that, without adjustable parameter, a stretched exponential
describes adequately our experimental results for the distribution of
acceleration fluctuations. Although not shown in the Figure, we verified that
this result holds also for the data shown in panels (b) and (c), i.e. for $A =
10$~mm and $A=15$~mm, respectively. This non-gaussianity of the acceleration
components is a manifestation of intermittency in turbulence, which is known to
have direct consequences on the forces exerted by the flow on an inertial
particle \cite{21}. 


\section{Conclusions}

An instrumented particle for the characterization of turbulent flows is
presented in this study. This device takes the form of a 36~mm sphere.  It
constitutes an autonomous local measurement device capable of measuring both its
translational acceleration and angular velocity components employing a high
acquisition rate (up to values of the order of the kHz), as well as recording
them on an embarked microSD removeable memory card. A LiPo battery powers the
electronics, providing the particle with up to 8~hours of autonomous operation.
Its sensors have a resolution of 16~bits, and maximum acquisition frequencies of
1~kHz for the accelerometer and 8~kHz for the gyroscope; as well as various
user-programmable dynamic ranges. This instrumented particle has a total mass of
slightly less than 20~g, resulting in an effective mass density of 819~kg/m$^3$.
Moreover, this density can be increased (within certain limits) by the inclusion
of properly dimensioned ballasts inside the sphere. 

It is worth pointing out that all the IP's electronic components are COTS
(commercial off-the-shelf), thereby presenting considerable advantages, such as
easy accessibility, low comparative cost, seamless upgrading and open
documentation. 

The particle's design and its properties, ranging from the choice of electronic
components to the construction of the casing and its assembly, is discussed in
detail. A calibration protocol is established and the IP's sensors are validated 
through two controlled experiments. For the 3-axis accelerometer, in particular,
this validation stage shows very good agreement between the signals measured
by the IP, those obtained by a particle tracking technique and a theoretical
prediction. 

Finally, the last Section presents an application of the IP to the
statistical characterization of a turbulent flow.
For this purpose we choose a surface wave turbulence scenario, generated 
in a water-filled wave tank by the stirring motion of piston-type 
wavemakers. A steady turbulent wave field of controlled characteristics is 
created by forcing the wavemakers with a white-noise random signal of amplitude
$A$ within the 0--4~Hz frequency band. In order to explore the effect of the
energy injection on the flow, four different forcing amplitudes are considered. 

This turbulent flow is then studied simultaneously with our instrumented
particle (two identical IPs are used in this case, actually) and their movement
is also registered by a 3D-PTV system for the purposes of comparison. 

Both measurement methods (IP and PTV) resulted in similar translational
acceleration statistical behavior, for all axes and forcing intensities. The
statistics of translational acceleration fluctuations measured by the IP 
as a function of the forcing amplitude reveal the departure from gaussianity
and the emergence of strongly non-gaussian distributions,
indicated by heavy tails beyond three standard
deviations and sub-normal behavior around the mean value. These results are in
qualitative agreement with previous studies carried out in other turbulent
systems. Additionally, it is shown that, by using a stretched exponential
distribution (with no adjustable parameters) it is possible to adequately
describe the non-gaussian behavior observed for the acceleration PDFs.
The non-gaussianity of these distributions accounts for the intermittency in the
flow, and the action of forces much higher than the average with occurrence
rates above what is expected for a normal distribution.  Additionally, the
angular velocity of the particles was studied. Temporal angular velocity
measurements coming solely from the instrumented particle present, in 
contrast, gaussian distributed fluctuations. 

Beyond the particularities inherent to this chosen case study, 
the results drawn from the application of the IP to a free-surface
turbulent flow and, more importantly, the quality of their agreement with 
their PTV counterparts and with previous works in similar physical scenarios,
constitute a proof of both the feasibility and potentiality of using the IP for
the characterization of particle dynamics in such flows. 


\begin{acknowledgements}
The authors acknowledge financial support from UBACYT Grant No. 20020170100508BA 
and PICT Grant No. 2015-3530.
\end{acknowledgements}

 \bibliographystyle{unsrt}       
\bibliography{references}

\end{document}